\documentclass[twocolumn]{aastex631}

\begin{document}

\newcommand{\Fe}{\ion{Fe}{1}~6302~\AA}
\newcommand{\NaD}{\ion{Na}{1}~D1~5896~\AA}
\newcommand{\ca}{\ion{Ca}{2}~8542~\AA}

\newcommand{\Blos}{\ensuremath{B_\mathrm{LOS}}}



\title{Multi-line Spectropolarimetric Observation of Flare Ribbon Fine Structures with ViSP/DKIST}


\author{Rahul Yadav}
\affiliation{Laboratory for Atmospheric and Space Physics, University of Colorado, Boulder, CO 80303, USA; \textnormal{rahul.yadav@lasp.colorado.edu}}

\author{Maria D. Kazachenko}
\affiliation{Laboratory for Atmospheric and Space Physics, University of Colorado, Boulder, CO 80303, USA; \textnormal{rahul.yadav@lasp.colorado.edu}}
\affiliation{ National Solar Observatory, 3665 Discovery Drive, 80303, Boulder, CO, USA}
\affiliation{Dept. of Astrophysical and Planetary Sciences, University of Colorado, Boulder, 2000 Colorado Ave, 80305, Boulder, CO, USA}

         
\author{Gianna Cauzzi}
\affiliation{ National Solar Observatory, 3665 Discovery Drive, 80303, Boulder, CO, USA}

\author{Cole A Tamburri}
\affiliation{DKIST Ambassador}
\affiliation{ National Solar Observatory, 3665 Discovery Drive, 80303, Boulder, CO, USA}

\author{Marcel Corchado}
\affiliation{ National Solar Observatory, 3665 Discovery Drive, 80303, Boulder, CO, USA}

\author{Ryan French}
\affiliation{Laboratory for Atmospheric and Space Physics, University of Colorado, Boulder, CO 80303, USA; \textnormal{rahul.yadav@lasp.colorado.edu}}

\begin{abstract}
We present an analysis of flare ribbon fine structure observed during a GOES C2-class flare using high spatial and spectral resolution multi-line spectropolarimetric observations from the ViSP instrument at the DKIST. The ViSP recorded full Stokes spectra in three lines: \ion{Fe}{1} 6301~\AA~line pair, \ion{Na}{1}~D1, and \ion{Ca}{2} 8542~\AA. To infer the stratification of temperature and line-of-sight (LOS) velocity across the ribbon, we performed non-LTE multi-line inversions of ViSP spectra.
In the red wing of the \ion{Ca}{2} line, we identified multiple compact, roundish, and quasi-equally spaced bright structures, referred to as ribbon blobs, embedded within the flare ribbon. The sizes of these blobs range from 320 to 455~km, and they are spaced roughly $\sim$1100~km apart. These features exhibit complex spectral profiles with pronounced asymmetries and double peaks near the line core of the \ion{Ca}{2} 8542 line. The blob regions were found to be significantly hotter (by $\sim$1 kK) at $\log~\tau_{500} = -4$ compared to surrounding ribbon areas ($\sim$7~kK). The LOS velocity maps revealed both upflows and downflows at $\log~\tau_{500} = -4$ and $\log~\tau_{500} = -3$, respectively. We discuss the plausible origins of these fine structures in the chromosphere, which may be related to electron beam heating, plasma draining, or tearing-mode instabilities in the reconnecting current sheet. 
 \end{abstract}

\section{Introduction}
Solar flares are among the most energetic phenomena in the solar atmosphere, releasing vast amounts of energy within minutes. One of the most prominent features of flares observed in the lower atmosphere is the appearance of flare ribbons, spatially extended brightenings that typically appear in pairs and outline the footpoints of newly reconnected magnetic field lines \citep{2011SSRv..159...19F, 2022SoPh..297...59K}. These ribbons are the chromospheric and transition regions signatures of energy deposition from the reconnection site, transported via energetic particles, thermal conduction and/or Alfv\'en waves \citep{2015ApJ...807L..22G, 2016ApJ...827..101K}.


While standard models \citep{1964NASSP..50..451C, 1966Natur.211..695S, 1974SoPh...34..323H, 1976SoPh...50...85K} provide a broad framework for understanding flare morphology and energetics, they do not explain the origin of the dynamic and fine-scale structures observed within flare ribbons by high-resolution telescopes \citep{2014ApJ...788L..18S,2015ApJ...810....4B, 2025A&A...693A...8T}. 

The advent of high-resolution ground-based solar telescopes---such as the 1-m Swedish Solar Telescope (SST; \citealt{2003SPIE.4853..341S}), the 1-m New Vacuum Solar Telescope (NVST; \citealt{2014RAA....14..705L}), the 1.5-m GREGOR telescope \citep{2012AN....333..796S}, the 1.6-m Goode Solar Telescope (GST; \citealt{2012SPIE.8444E..03G}), and most recently, the 4-m National Science Foundation's Daniel K. Inouye Solar Telescope (DKIST; \citealt{2020SoPh..295..172R})---has made it possible to resolve and study subarcsecond structures and their evolution in the lower solar atmosphere.

High-resolution observations from these facilities have revealed that flare ribbons often exhibit complex, fine-scale features, including narrow, wavy, or dot-like bright substructures, commonly referred to as ribbon blobs, plasma blobs, bright knots, flare kernels, ribbon fronts or riblets \citep{2014ApJ...788L..18S,  2016NatSR...624319J,French2021, 2023ApJ...944..104P, 2024A&A...685A.137P, 2024ApJ...965...16C, 2025A&A...693A...8T,Lorincik2025, 2025arXiv250701169S}. 

These features are often observed in the red wings of chromospheric lines. For example, \citet{2014ApJ...788L..18S} reported numerous sub-arcsecond bright knots in the red wing of H$\alpha$ with widths of $\sim$100~km. More recently, using SST observations \citet{2025A&A...693A...8T} identified circular, spatially periodic blob-like structures with sizes of 140--200~km in the red wings of \ion{Ca}{2}~8542, H$\alpha$, and H$\beta$. 

Although several models have been proposed to explain these fine structures, their physical origin remains unclear. \citet{2014ApJ...788L..18S} suggested that the bright blobs may be powered by Joule heating. Recent high-resolution three-dimensional MHD simulations \citep{2021ApJ...920..102W, 2025arXiv250400913D} have proposed that spiral or wave-like ribbon fine structures could instead be signatures of tearing-mode instability in the flare current sheet. 

The study of ribbon fine structures remains of great importance, as they are signatures of magnetic reconnection in the corona, and could potentially constrain the electron energy flux required for flare modeling \citep{2017NatCo...815905D, 2016ApJ...827..101K, 2021ApJ...920..102W,2024ApJ...965...16C,2024ApJ...970...21K,2025arXiv250400913D}. However, such observations remain rare and are mostly limited to ground-based facilities operating under optimal seeing conditions.

In this study, we present an analysis of flare ribbon fine structures using high spatial and spectral resolution and multi-line spectropolarimetric observations from the Visible Spectro-Polarimeter (ViSP; \citealt{2022SoPh..297...22D}) at the DKIST. 
Section~\ref{Sec:observations} outlines the observations and data reduction. Methodology and results are presented in Sections~\ref{sec:method_and_data} and \ref{sec:results}, followed by discussion in Section~\ref{sec:Discussion} and conclusions in Section~\ref{sec:conclusion}.
\begin{figure*}
    \centering
    \includegraphics[clip,trim=0.0cm 3.2cm 0.0cm 1cm,width=1\linewidth]{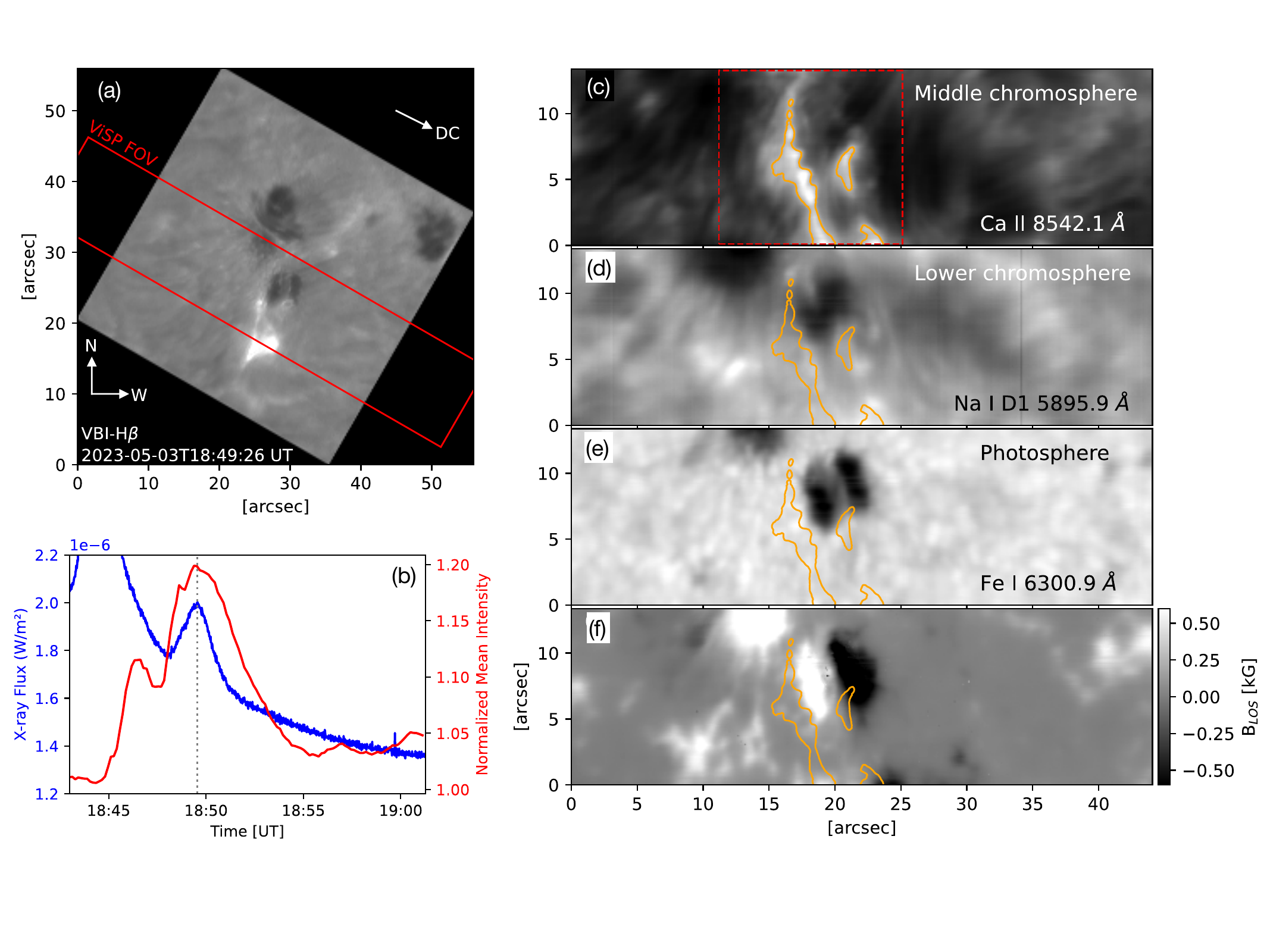}
    \caption{Overview of the C2-class flare (SOL2023-05-03T18:49) in AR 13293. (a) VBI FOV showing flare ribbons and the ViSP FOV. (b) Temporal evolution of GOES X-ray flux and mean VBI/H$\beta$ intensity are shown in blue and red colors, respectively. The dashed vertical line depicts flare peak time. (c) line core intensity map in arm \ion{Ca}{2}~8542~\AA, (d) line core intensity map in arm \ion{Na}{1}~D1~5896~\AA, (e) continuum intensity map, and (f) line-of-sight magnetic field derived from \ion{Fe}{1}~6302~\AA. The dashed red box highlights FOV shown in Figure~\ref{fig:ribbon_structure_overview}.}
    \label{fig:vbi-overview}
\end{figure*}

\section{Data Overview}
\label{Sec:observations}
\subsection{Data Acquisition}

In the following, we describe the summary of DKIST data. On 3 May 2023, from 18:43 to 21:27 UT, the ViSP and Visible Broadband Imager (VBI; \citealt{2021SoPh..296..145W}) instruments at DKIST observed active region NOAA 13293. The center of the DKIST field of view (FOV) was positioned at helioprojective coordinates X = $-580$\arcsec\ and Y = $255$\arcsec\ (N13E37). The data acquisition and reduction procedures for this dataset are described in detail by \cite{2024ApJ...973L..10Y}. Below, we provide a brief summary of the dataset.
The ViSP simultaneously captured data from three distinct spectral regions using separate detectors: the \ion{Fe}{1} 6302~\AA\ line pair (hereafter, \ion{Fe}{1}) with arm 1, the \ion{Na}{1} D1 5896~\AA\ line  (hereafter, \ion{Na}{1}) with arm 2, and the \ion{Ca}{2} 8542~\AA\ line  (hereafter, \ion{Ca}{2}) with arm 3. For our observations, the width of the slit was set at 0.107\arcsec. Since the pixel scale along the slit was finer than the scan step size for all arms, we rebinned the pixels along the slit to improve the signal-to-noise ratio and to obtain square pixels of 0.107\arcsec. Spatial scanning proceeded perpendicular to the slit direction in 125 discrete steps, resulting in an overall FOV of 13.4\arcsec~in width by the length of the slit. The lengths of the slit varied across the three arms: 75.5\arcsec\ for the \ion{Fe}{1} channel, 61\arcsec\ for the \ion{Na}{1} channel, and 49\arcsec\ for the \ion{Ca}{2} channel. Spectral pixel sampling for these channels are 12.8~m\AA, 14~m\AA, and 18.8~m\AA, respectively. By comparing quiet-Sun spectra observed at disk center with solar atlas profiles, we estimate the effective spectral resolving powers to be approximately 140,000 for arm1 (\ion{Fe}{1}), 131,000 for arm2 (\ion{Na}{1}), and 142,000 for arm3 (\ion{Ca}{2}).
At each scan step, full-Stokes polarization data were collected using four modulation cycles, each comprising ten modulation states. This setup led to an integration time of 1.5 seconds per slit position, producing a total of 48 raster maps with a temporal cadence of 3.11 minutes.

Alongside ViSP, the VBI captured high-resolution intensity images in the H$\beta$ line at a spatial sampling of 0.01\arcsec~per pixel and a cadence of 9 seconds. The speckle-reconstructed VBI data were provided by the DKIST
Data Center.  

\subsection{Overview of Observations}
The AR NOAA 13293 produced multiple flares within the DKIST observing time. In this study, we focus on a small C2-class flare captured by both VBI and ViSP. Figure~\ref{fig:vbi-overview} shows an overview of this flare in the H$\beta$ channel of the VBI and ViSP spectra. This short-lived flare started, peaked and ended at 18:48:00, 18:49:33, and 18:50:05 UT, respectively. As shown in Figure~\ref{fig:vbi-overview}, the GOES X-ray profile of this flare occurred during the decay phase of a C2.7 flare 
developing in a different AR (NOAA 13296). Therefore, the actual GOES class of the observed flare may be lower than the C2-class. The ViSP captured most of the flare ribbon extent, except for the southern ribbon structure. Figure~\ref{fig:vbi-overview} also shows the appearance of the ribbon 
in the \ion{Ca}{2}, along with \ion{Na}{1} and \ion{Fe}{1} spectral lines.

    


\begin{figure}[!ht]
    \centering
    \includegraphics[clip,trim=0.7cm 0.cm 0.4cm 0.cm,width=0.5\textwidth]{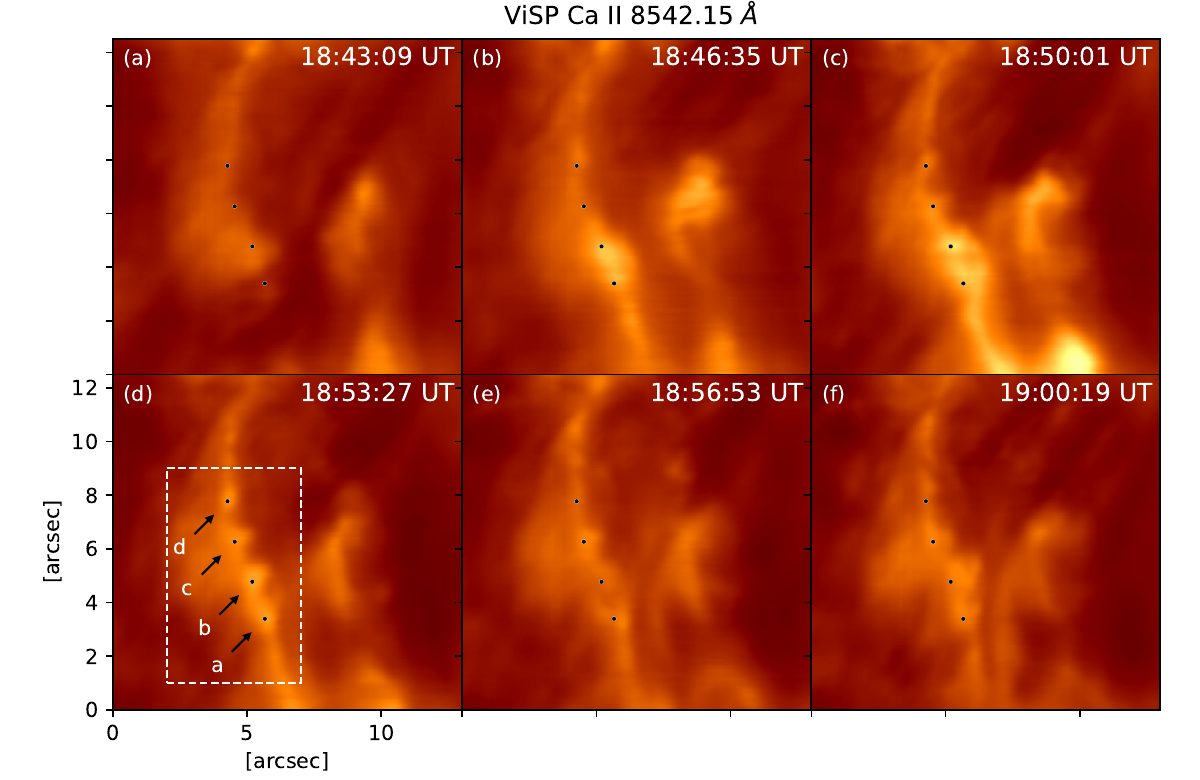}
    \vspace{4mm}
    \includegraphics[clip,trim=0.4cm 0.2cm 0.2cm 0.2cm,width=0.5\textwidth]{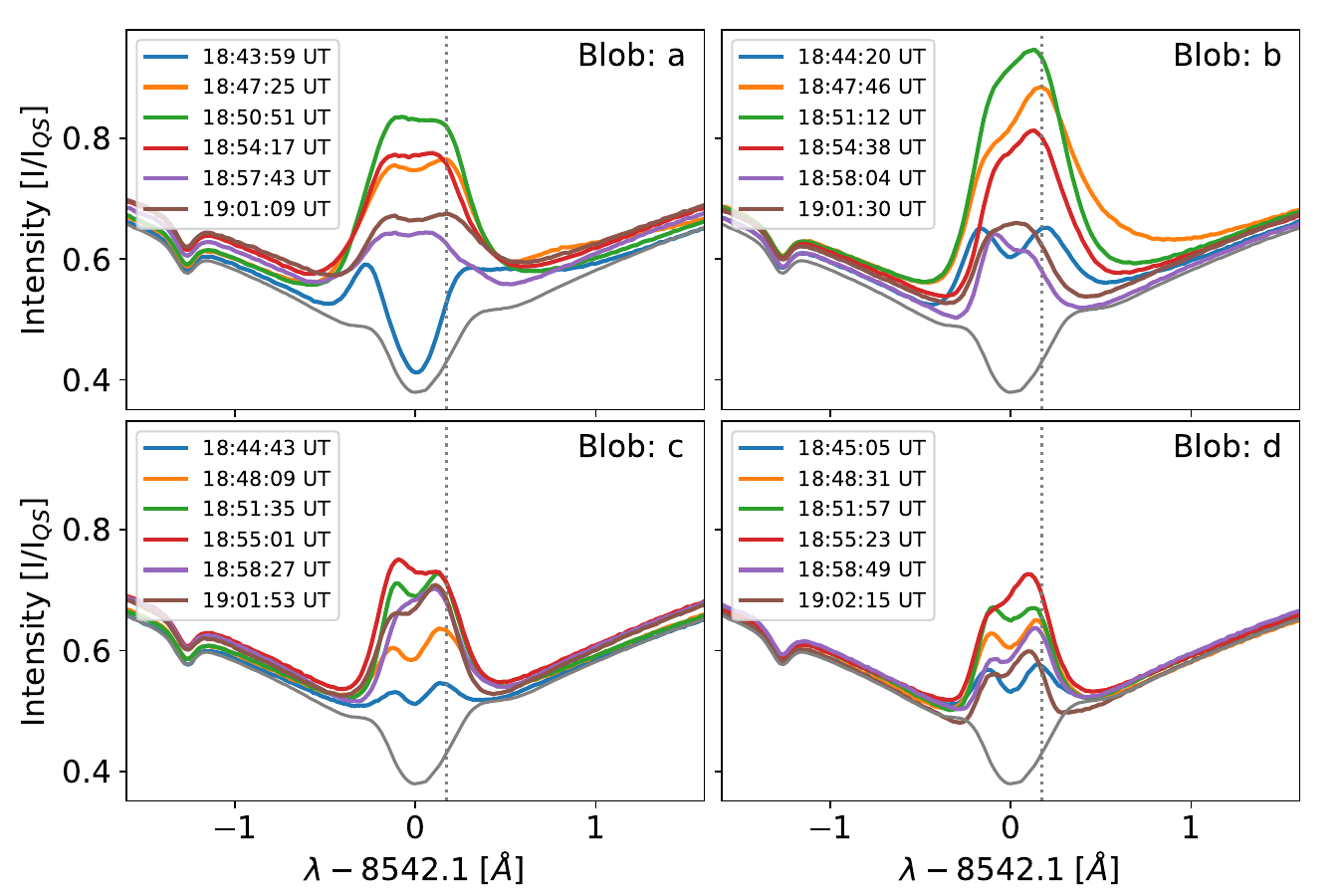}
    \vspace{-8mm}
    \caption{Ribbon fine-structures observed by ViSP in the red box highlighted in Figure~\ref{fig:vbi-overview}c. Panels (a--f) show temporal evolution of ribbons in the red-wings of \ion{Ca}{2} spectra (note that the ViSP raster scans proceed from bottom to top of the panels; the reported time corresponds to the first slit position in the scan). Black dots in panels show the location of the center of the bright blobs, as identified at the time of panel d; the dashed white box in panel d is reproduced in Fig. \ref{fig:k-mean_profiles}.  Bottom panels show temporal evolution of Ca II profiles at the marked pixels in blobs a, b, c and d. The gray line refers to the mean quiet-Sun spectra taken from regions away from the ribbons.. The dashed vertical line marks the location of monochromatic images shown in panels (a--f).}
\label{fig:ribbon_structure_overview}
\end{figure}

\begin{figure}[!ht]
    \centering
    \includegraphics[width=0.5\textwidth]{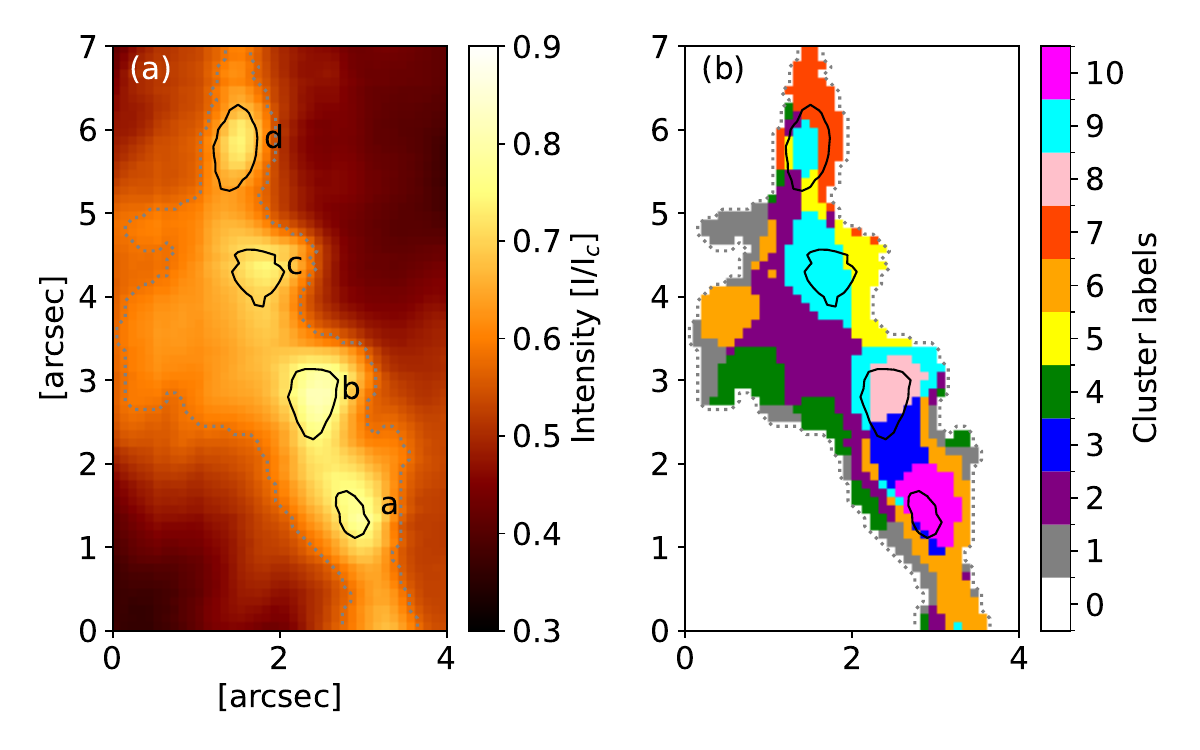}
    \includegraphics[width=0.5\textwidth]{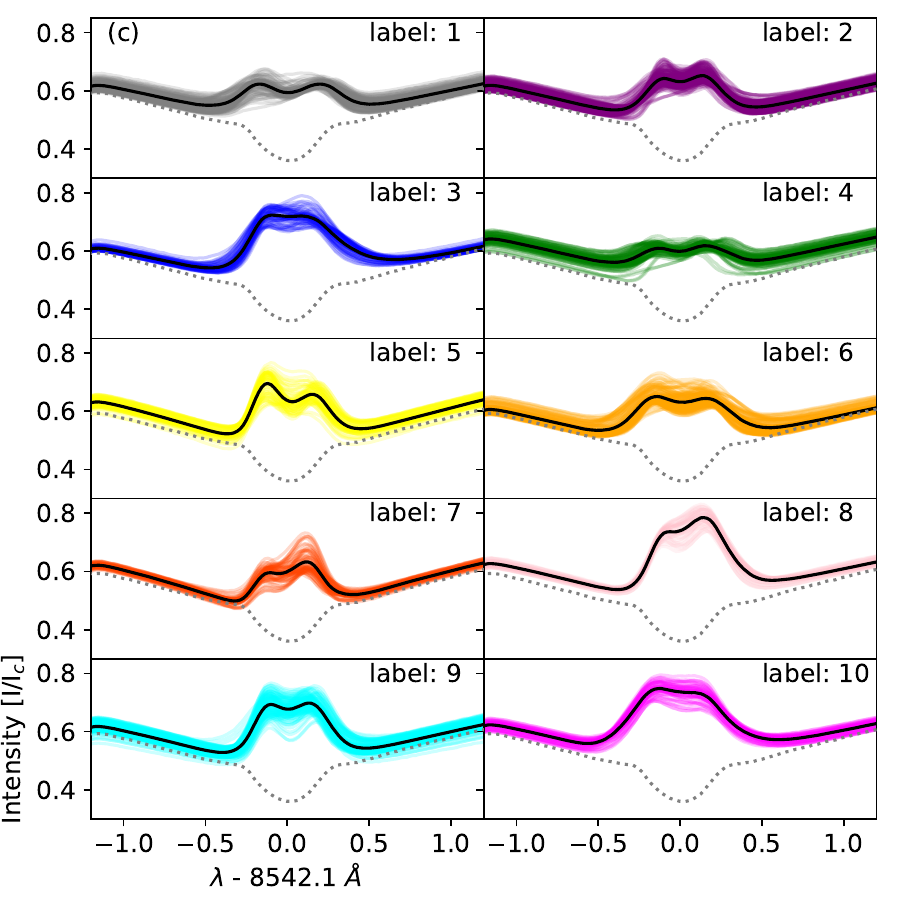}
    
    \caption{\textit{K}-means clustering of \ion{Ca}{2} spectral profiles for the panel d shown in Figure~\ref{fig:ribbon_structure_overview}. (a) Intensity map of the blobs within the white box in Figure~\ref{fig:ribbon_structure_overview}d. (b) Grouping of spectral profiles, labeled from 1 to 10, shown in distinct colors. The dotted and solid contours mark the \textit{K}-means clustering and the ribbon blobs locations, respectively. (c): Clustered \ion{Ca}{2} profiles, the solid black lines represent the mean spectra for each group, and the dotted gray lines show the corresponding quiet-Sun spectra.} 
    \label{fig:k-mean_profiles}
\end{figure}
\section{Methods and Data 
Analysis}\label{sec:method_and_data}
\subsection{Inversion of the Spectropolarimetric Data}
We used a single ViSP \ion{Fe}{1} line (e.g., \ion{Fe}{1}~6302.5\AA) and the SPIN Milne–Eddington inversion code \citep{2017SoPh..292..105Y} to retrieve the magnetic field vector in the photosphere. To derive the stratification of physical parameters---such as temperature, magnetic field, line-of-sight (LOS) velocity, and microturbulent velocity---within the ribbon FOV, we employed the multi-line non-LTE STiC inversion code \citep{2019A&A...623A..74D}. STiC is built around a modified version of the RH code \citep{Uitenbroek2001}, which computes atomic populations assuming statistical equilibrium and a plane-parallel geometry. The equation of state is adopted from the Spectroscopy Made Easy (SME) code described in \cite{2017A&A...597A..16P}, and the radiative transfer equation is solved using cubic Bézier solvers \citep{2013ApJ...764...33D}.

We used all Stokes profiles of the \ion{Ca}{2}, \ion{Na}{1}, and \ion{Fe}{1} lines for the inversion with the STiC code. These lines are sensitive to different layers of the solar atmosphere: \ion{Fe}{1} to the photosphere, \ion{Na}{1} to the lower chromosphere, and \ion{Ca}{2} to the middle chromosphere \citep{2017SSRv..210..109D}. Consequently, the use of multiple lines enables a more accurate reconstruction of atmospheric stratification than would be possible using a single line alone. 

The \ion{Fe}{1} line pair was treated under the assumption of LTE conditions, whereas the \ion{Na}{1} and \ion{Ca}{2} in Non-LTE.
To accelerate the STiC inversion, we used a machine-learning-trained model atmosphere as the initial guess. This approach significantly improves both the speed and the accuracy of inversion \citep{2021A&A...649A.106Y}. We note that STiC inversions are performed under the assumption of hydrostatic equilibrium, which may not be ideal for flaring regions. Nevertheless, this approximation has been used by several authors 
\citep{Kuridze2018, 2019A&A...621A..35L, 2021A&A...649A.106Y, 2023FrASS..1033429S, Kuckein2025},
and represents a reasonable approximation for flaring pixel after the (brief) impulsive phase.

\subsection{\textit{K}-means Clustering of Profiles}
\label{subsec:kmean}
We apply \textit{K}-means algorithm \citep{macqueen1967} to the observed \ion{Ca}{2} profiles to identify and classify similar profiles within the flare ribbon area. The \textit{K}-means clustering is  an unsupervised machine learning technique, which is widely used to classify the spectral profiles \citep{2011A&A...530A..14V,2018ApJ...861...62P,2019ApJ...875L..18S}. The \textit{K}-means partitions the \textit{N} observed profiles into \textit{K} distinct clusters or labels, defined by their respective centroids. 
The algorithm operates iteratively, beginning with the random initialization of \textit{K} centroids. It then iterates between two steps: assigning each data point to the nearest centroid, and updating the centroids as the mean of all points assigned to each cluster. This process continues until convergence, typically when the assignments no longer change or the centroids stabilize.
For a detailed description of the \textit{K}-means procedure and its limitations, we refer the reader to \cite{2023LRSP...20....4A}.

\section{Results}
\label{sec:results}
\subsection{Flare Ribbon Fine Structure Observed by ViSP}
Flare ribbons associated with the C2-class flare are observed in both the VBI and ViSP images. Although the VBI offers higher spatial and temporal resolution compared to ViSP, fine ribbon substructures are not evident in the VBI images. This 
is probably due to the spectral smearing of ribbon features resulting from the broadband nature of VBI imaging across H$\beta$ (as mentioned in the Introduction, these features are most often observed only in the red wing of chromospheric lines). In contrast, the high spectral resolution ViSP data reveal fine ribbon substructures in the red wings of the \ion{Ca}{2} line. These structures, best identified around the peak time of the flare,  appear compact, roundish, and quasi-equally spaced; they are labeled as a, b, c, and d in Figure~\ref{fig:ribbon_structure_overview}d. We refer to them as \textit{ribbon blobs}. These blobs are approximately 1.4 times brighter than the surrounding ribbon regions and are clearly discernible (see Figure~\ref{fig:k-mean_profiles}a). 

We manually identify the blob locations using an intensity threshold and determine their mean sizes by fitting ellipses to their boundaries. The obtained semi-minor axes (A) for all blobs range from 0.16 to 0.26 arcsec, while the semi-major axes (B) range from 0.23 to 0.43 arcsec. The ellipticity of the blobs ranges from 0.55 to 0.95, indicating round to elliptical shapes. We used the A and B axes of the fitted ellipses to estimate the equivalent circular diameters, calculated as 2$\sqrt{AB}$ by equating the areas of the ellipse and the corresponding circle. The resulting estimated blob widths range from 0.45\arcsec\ to 0.63\arcsec\ (320–455 km), with blobs a and d being the smallest and largest, respectively (see Figure~\ref{fig:k-mean_profiles}a)
The separations between consecutive blobs, measured from the centroid of the ellipses, range from 1.4\arcsec~to 1.7\arcsec~(1015–1230 km).

ViSP observed the ribbon region by scanning the slit across the FOV with a step size of 0.107\arcsec (78~km). Each raster scan took approximately 3.11 minutes to cover the full FOV, which could potentially introduce artifacts in the appearance of fine structures if they evolve very rapidly. However, as shown in Figure~\ref{fig:ribbon_structure_overview} (top row), we find that the small-scale blob-like structures
can be identified in 2-3 consecutive rasters 
during both the peak and early decay phases of the flare. This suggests that the ribbon blobs have lifetimes of several minutes, and might be related to sites of enhanced heating within the chromospheric ribbon. 

Recently, similar fine structures in flare ribbons have been reported by \citet{2025A&A...693A...8T} using high spatial and temporal resolution H$\beta$ observations obtained with the SST. Additionally, \citet{2014ApJ...788L..18S} identified numerous small-scale ($\sim$100~km) `bright knots' in a long, thread-like ribbon structure using high-resolution observations from the New Solar Telescope (now known as the GST) in the red wing of the H$\alpha$ line in a C2.1-class flare.

The detection of these fine structures in different chromospheric lines (e.g., H$\beta$, H$\alpha$, and \ion{Ca}{2}) suggests that they may form over a large portion of the chromosphere.

\begin{figure*}[]
    \centering
    \includegraphics[clip,trim=0cm 0.cm 0.cm 0.cm,width=1\textwidth]{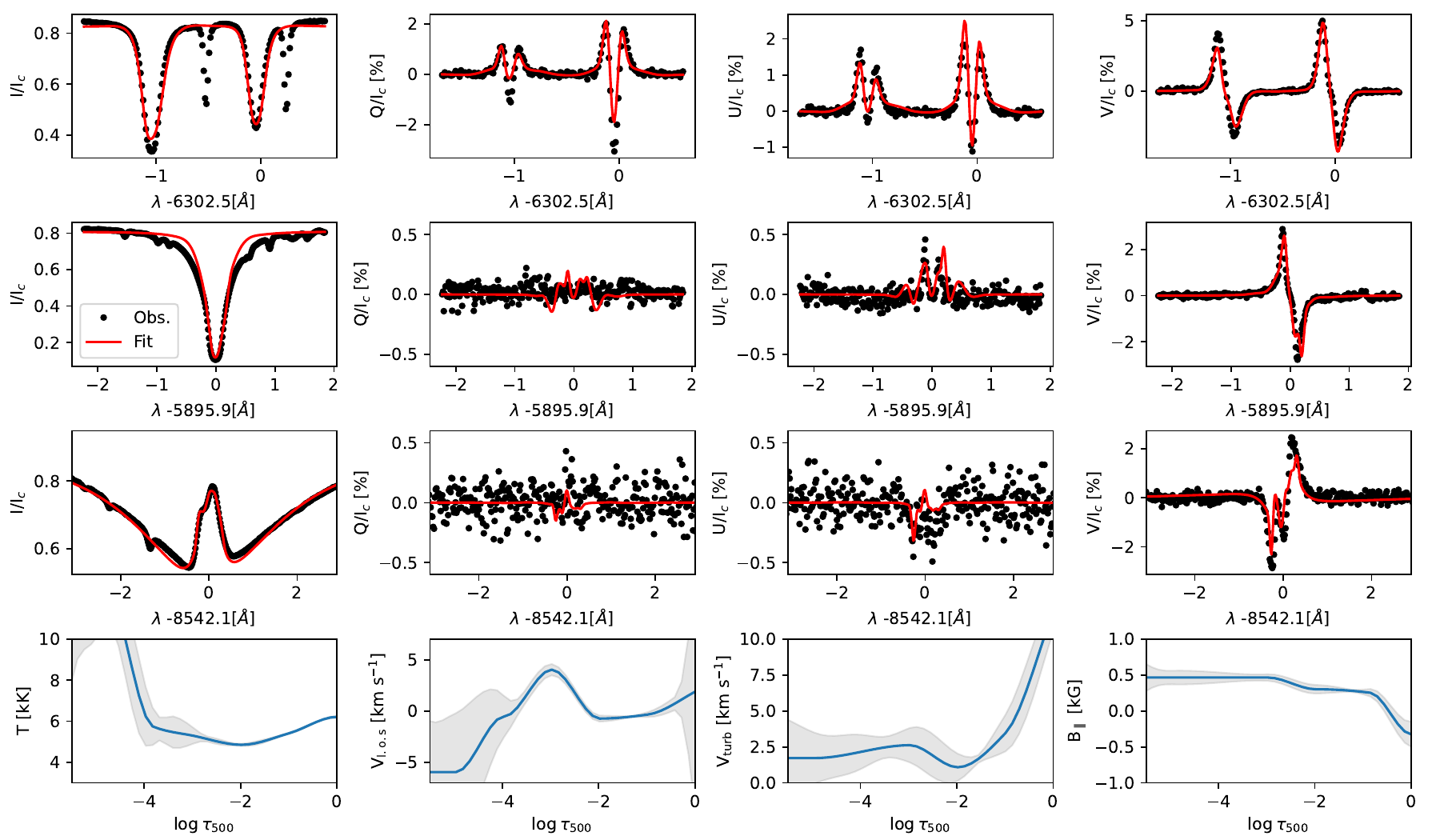}
    \caption{Multi-line inversion of a pixel (marked by dot) located in blob b shown in Figure~\ref{fig:ribbon_structure_overview}d. Top three rows: observed (dotted line) and best fits (solid red line) of the
Stokes profiles. In each row all Stokes profiles are shown for the \ion{Fe}{1}, \ion{Na}{1}~D1, and \ion{Ca}{2} spectral lines. Bottom row: inferred model atmosphere---temperature (T), LOS velocity (V$_{LOS}$), micro-turbulent velocity (V$_{turb}$) and LOS magnetic field (B$_{||}$). The gray shaded regions refers to the uncertainties in the inferred parameters.}
    \label{fig:stic_fit}
\end{figure*}
\subsection{\textit{K}-means Clustering of Ribbon Profiles}
The high-spectral resolution of ViSP spectra in the \ion{Ca}{2} allows us to detect complex profiles within the ribbons. As shown in Figure~\ref{fig:ribbon_structure_overview}, the temporal evolution of the \ion{Ca}{2} profiles at the blob locations exhibits pronounced asymmetries and spectral line broadening. Additionally, all profiles display two peaks near the line center of \ion{Ca}{2}. To investigate how these profiles are distributed within the blobs and the surrounding ribbon area, we applied the \textit{K}-means clustering algorithm (see Sect.\ref{subsec:kmean}) to a single ViSP map at 18:53 UT  (Figure~\ref{fig:ribbon_structure_overview}d), where the blobs are best identified.

Figure~\ref{fig:k-mean_profiles} shows the spatial distribution of the \ion{Ca}{2} profiles grouped using the \textit{K}-means method. After testing several cluster numbers, we found that the blob-associated profiles are best grouped using ten clusters. The clustering analysis reveals that the ribbon profiles all show a self-reversed  emission and are notably broadened with respect to the quiet Sun, as often observed.  Notably, the profiles associated with the blobs (clusters labeled 8, 9 and 10) are significantly more intense than those in the surrounding ribbon region. 

Many profiles within the ribbon area are also asymmetric (e.g., labels 5, 7 and 8 in Figure~\ref{fig:k-mean_profiles}), with one of the peaks around line core sensibly higher than the other. While the profiles located within the ribbon at different positions appear similar in some cases (see, e.g., labels 1 and 6 in Figure~\ref{fig:k-mean_profiles}), their amplitudes differ slightly. These similar-looking profiles could be due to comparable atmospheric conditions at their respective locations, which shape their overall structure. The model atmosphere inferred from the ViSP spectra within the ribbon is discussed in more detail in Sect.~\ref{sec:stratification}.

Contrary to our findings, \citet{2025A&A...693A...8T} did not find double peaks near the line core in the \ion{Ca}{2} profiles observed with the CRISP instrument at the SST. This could be due to different flaring conditions (we note their \ion{Ca}{2} intensity is sensibly higher than what shown in Figure \ref{fig:ribbon_structure_overview}). We also note that the ViSP spectral resolution is roughly two times better than the CRISP/SST. This higher spectral resolution combined with a cadence of 1.5s per slit, enabled the detection of double peaks in the blob and nearby ribbon pixels. For comparison, the above mentioned SST \ion{Ca}{2} profiles, including two more spectral lines, have a cadence of 40~s. We note that the spectral sampling for the single \ion{Ca}{2} line may be lower than the cadence of their observations. Given the highly dynamic nature of the profiles during flaring, such a cadence may introduce smearing effects.

The presence of asymmetry in the ViSP spectra indicates complex velocity structures or multiple atmospheric components along the line of sight, which could be unresolved in more sparsely sampled or low-cadence data. 

\begin{figure*}[]
    \centering
    \includegraphics[clip,trim=1.2cm 2.2cm 2.2cm 2.cm,width=0.85\textwidth]{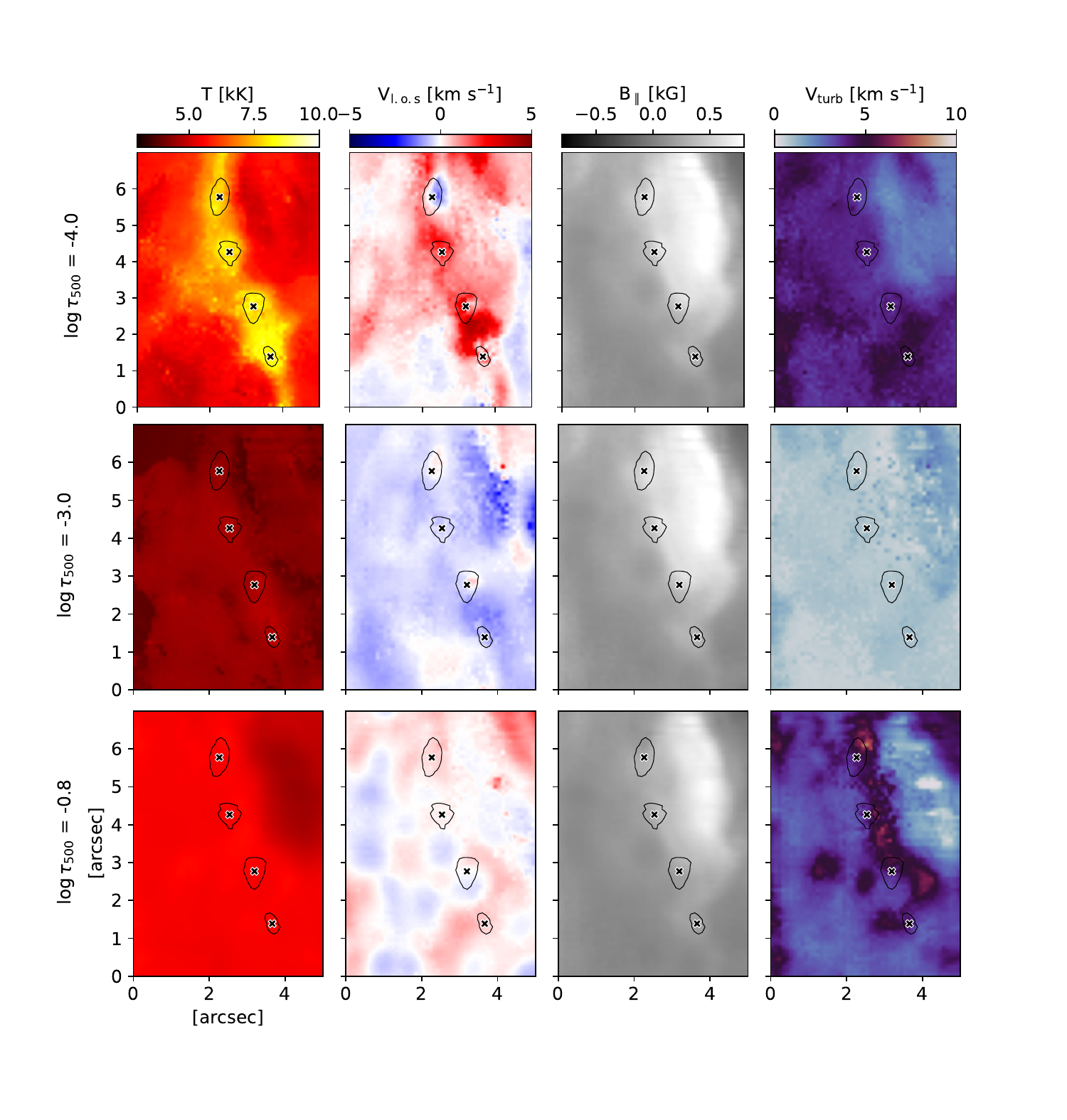}
\caption{Stratification of physical parameters inferred from the non-LTE inversion. Each row (from bottom to top) shows the parameters at different optical depths: $\log~\tau_{500} = -0.8$ (mid photosphere), $\log~\tau_{500} = -3$ (lower chromosphere), and $\log~\tau_{500} = -4$ (middle chromosphere). Left column: 
Temperature maps at the respective optical depths. Second, third, and fourth columns: Same as the left column, but for LOS velocity, LOS magnetic field, and microturbulent velocity, respectively. Black contours indicate the blob locations, and cross symbols mark their centers, as shown in Figure~\ref{fig:k-mean_profiles}a.}
    \label{fig:stic_strat}
\end{figure*}

\subsection{Stratification of Temperature and LOS velocity from Multi-line Inversion}
\label{sec:stratification}
To retrieve the stratified model atmosphere of the flare ribbon, we simultaneously inverted all three ViSP spectral lines (\ion{Fe}{1}, \ion{Na}{1}, and \ion{Ca}{2}) using the STiC code (see Sect.\ref{sec:method_and_data}). From the multi-line inversion, we derived the height-dependent stratification of physical parameters as a function of the logarithm of the optical depth at 500~nm (log $\tau_{500}$). We note that the \ion{Ca}{2} line has limited sensitivity below log $\tau_{500} \approx -4.5$ (higher chromospheric layer) \citep{2021A&A...649A.106Y,2025ApJ...985..157D}. At log $\tau_{500}$ smaller than this, the parameter are less reliable.

We inverted only one ViSP raster map (Figure~\ref{fig:ribbon_structure_overview}d), as the blobs are clearly visible in this scan. Figure~\ref{fig:stic_fit} shows an example of a multi-line inversion for a pixel located within blob~b. The synthetic profiles, generated from a single set of model parameters, exhibit excellent agreement with the observed profiles, demonstrating the consistency and reliability of the inversion. 

Figure~\ref{fig:stic_strat} presents the stratification of physical parameters in the ribbon region. The temperature stratification shows blobs as sites of enhanced heating within the ribbon,
particularly around $\log~\tau_{500} = -4$ (mid chromosphere). The mean temperature at the blob site reached $\sim$8~kK, whereas the near by ribbon pixels are $\sim$7~kK. This temperature range is in agreement with previous flare observations \citep{2017ApJ...846....9K, Kuridze2018,2021A&A...649A.106Y}. 

Self-reversed emission \ion{Ca}{2} profiles shown in Figure~\ref{fig:k-mean_profiles} may arise due to relatively strong heating at the blob sites, whereas similar but less intense profiles at the ribbon boundaries are likely a result of relatively weaker heating in the chromosphere.

The LOS velocity stratification reveals the presence of weakly blue-shifted components (upflows) near $\log~\tau_{500} = -3$ and red-shifted components (downflows) around $\log~\tau_{500} = -4$ in the chromosphere, except at the site of blobs b and d, where the flow directions are reversed. The coexistence of oppositely directed plasma flows (or gradient in velocity) at the blob locations may produce asymmetric profiles in the \ion{Ca}{2} spectral lines.

The chromospheric downflows shown in Fig. \ref{fig:stic_strat} are moderate, around 4-5~km~s$^{-1}$ within the ribbons and blobs, while the upflows in deeper layers remain below $-1$~km~s$^{-1}$. In contrast, \citet{2025A&A...693A...8T} reported that \ion{Ca}{2} spectral profiles in similar flaring blobs display an excessive red wing component at about 20~km~s$^{-1}$. This could be due to their observations being acquired at a different phase of the flare development. However, we also note that this velocity estimates were not derived through spectral line inversions or other approaches, so the reported high values may carry significant uncertainties.

The stratification of the LOS magnetic field shows that the blobs are located in regions of weak positive polarity ($\sim$250~G in the photosphere), near the boundary between the pore and the quiet Sun. The microturbulent velocity (average 4~km~s$^{-1}$) exhibits no clear variation in the areas of the ribbon (except blob a) at chromospheric height ($\log~\tau_{500} \approx -4$). 

As demonstrated in Figure~\ref{fig:stic_fit}, the linear polarization signals (Stokes Q and U) in the \ion{Na}{1}~D and \ion{Ca}{2} spectral lines are too weak to reliably infer the magnetic field vector within the ribbons.

The uncertainties in the inferred parameters shown in Figure~\ref{fig:stic_fit} are estimated by performing 5000 inversions of the same profile, each using randomly generated initial model parameters under the same inversion configuration.

\section{Discussion}
\label{sec:Discussion}
In this work, we have presented an analysis of fine structures within a flare ribbon observed during a GOES C2-class flare, using high-resolution data from the ViSP and VBI instruments at DKIST. ViSP carried out spectropolarimetric observations in three spectral lines---\ion{Fe}{1}~6301~\AA\ line pair, \ion{Na}{1}~D1 5896~\AA, and \ion{Ca}{2}~8542~\AA---sampling the lower solar atmosphere. In addition to this, VBI captured high-resolution intensity images of the flare in the H$\beta$ broadband channel.

The high spatial and spectral resolution of ViSP enabled us to resolve fine-scale structures and investigate the complex shapes of the \ion{Ca}{2} profiles within the flare ribbon shortly after the impulsive phase. In the red wing of the \ion{Ca}{2} line, we identified four small, roundish, and quasi-equally spaced bright features that are significantly brighter than the surrounding ribbon emission. We refer to these features as ribbon blobs. Their sizes range from 0.45\arcsec\ to 0.63\arcsec\ (320–455 km), and they are separated by distances of approximately 1.4\arcsec~to 1.7\arcsec~(1015-1230 km). 

In contrast to the findings of \citet{2025A&A...693A...8T}, our \textit{K}-means classification of the profiles revealed that most pixels within the ribbon exhibited double peaks near the line center of \ion{Ca}{2}. Moreover, the profiles at the blob locations were notably more intense and asymmetric compared to the pixels a few arcsec away in the surrounding ribbon region.

To the best of our knowledge, we present the first multi-line non-LTE inversion of ribbon fine-structures using the state-of-the-art STiC code. The high spectral and spatial resolution of the ViSP spectra in three lines enabled a more accurate inference of the stratification of temperature and LOS velocity.

Temperature stratification revealed that the blob sites are significantly heated, reaching temperatures of approximately 8~kK in the chromosphere, compared to nearby ribbon pixels, which show values around 7~kK.  The LOS velocity maps 
indicate the presence of both upflows (blueshifts) and downflows (redshifts) in the ribbon area, exhibiting complex dynamics of the plasma. At $\log~\tau_{500} = -4$, most of the ribbon region is redshifted. However, the blob sites exhibit relatively stronger downflows of up to $\sim$4~km~s$^{-1}$. In contrast, in the deeper layers at $\log~\tau_{500} = -3$, the flows are mainly weak upflows ($\sim$1~km~s$^{-1}$). The presence of both upflows and downflows in the ribbon areas, including the blobs, could have produced the observed asymmetries in the \ion{Ca}{2} spectral lines.


The physical mechanisms responsible for the appearance of these roundish, blob-like ribbon fine structures remain poorly understood. 
In line with the standard flare model,  these downflows could be ascribed to the so-called ``chromospheric condensation", the rapid compression of the lower atmosphere due to the response to a sudden release of a large amount of energy \citep{1984SoPh...93..105I,2020ApJ...895....6G}. This seems consistent with the enhanced heating observed at the blob locations, suggesting that these features may be sites of more intense energy deposition during the magnetic reconnection process.
The weak upflows observed in the deeper layers ($\log~\tau_{500} = -3$) may be explained by the mechanism proposed by \citet{2019A&A...621A..35L}, in which, during the later phase of a flare, chromospheric condensations compress the deeper chromosphere, generating a shock. As a result, the deeper layer is heated and rebounds, producing weak upflows.

Additionally, from high resolution observations, \citet{2014ApJ...788L..18S} suggested that the observed bright blob-like structures in flare ribbons may be powered by Joule heating. However, we were unable to investigate the role of electric currents in our study due to the weak magnetic field strength at the ribbon site.

Flare models typically propose that electron beams originating from the coronal reconnection site bombard the chromosphere, producing bright ribbons and driving strong chromospheric upflows and downflows \citep{2016ApJ...827..101K, 2017NatCo...815905D, 2023ApJ...944..104P}. Such localized energy deposition is often associated with hard X-ray footpoint sources, which trace regions of intense electron precipitation \citep{2011ApJ...739...96K,2016ApJ...827...38R}. 

Recent short-exposure images (174~\AA) from the EUI/Solar Orbiter have revealed bright substructures in flaring ribbons that spatially coincide with the hard X-ray emission observed by the STIX instrument \citep{2024A&A...692A.176C}. This suggests that the brightest parts of the ribbons may be associated with a stronger electron energy flux.

Based on our observations, we speculate that the temperature enhancement in the blobs could result from a higher electron beam energy or flux compared to the surrounding ribbon regions. To test this hypothesis, parametric studies using various electron beam models are needed (e.g., \citealt{2022A&A...665A..50Y, 2023ApJ...944..104P}). Such investigations are now feasible with the RADYN simulation framework \citep{1997ApJ...481..500C, 2015ApJ...809..104A}, though they are beyond the scope of the present study.

The fine structures of flare ribbons have been explored through both analytical models and MHD simulations.
\citet{2021ApJ...920..102W} reported periodic spiral and wave-like features in flare ribbons, attributed to tearing-mode instabilities in the flare current sheet. The blobs identified in our observations are similar to the equidistant spiral features predicted by their analytical model. Recently, using high-resolution three-dimensional MHD simulations of an eruptive flare, 
\citet{2025arXiv250400913D} showed that plasmoids can imprint transient spiral features at the ribbon front, qualitatively resembling the ribbon blobs observed in our study, though their simulated structures are two orders of magnitude larger. Our results provide strong observational evidence supporting the occurrence of current sheet tearing during flare-associated magnetic reconnection.
 
Previous theoretical and numerical studies suggest that the fragmentation of current sheets into magnetic islands (plasmoids) due to tearing-mode instability can lead to intermittent and spatially localized energy release. Such processes are believed to play a critical role in particle acceleration and energy transport during solar flares, and can shape the fine structure of ribbons (e.g., \citealt{2006Natur.443..553D,2009PhPl...16k2102B,2010PhPl...17f2104H,2016ApJ...820...60G,2018ApJ...866....4L, 2024ApJ...970...21K, 2025arXiv250400913D}).

However, our ViSP observations are not well suited for investigating the lifetime and spatial motion of these flare fine structures. To explore their dynamics and test the above hypothesis in the lower solar atmosphere, observations with higher cadence and spatial resolution are needed. Further observations with ViSP in different configurations (e.g., spectroscopy-only mode) could help address these limitations. The recently commissioned DL-NIRSP Integral Field Unit spectrograph at DKIST \citep{2022SoPh..297..137J} also holds promise of fast, high spatial and spectral resolution spectroscopy, albeit over a small FOV.  These capabilities will be further enhanced by future state-of-the-art instruments at the 4-m European Solar Telescope (EST; \citealt{2022A&A...666A..21Q}) and the 2-m National Large Solar Telescope (NLST; \citealt{2010IAUS..264..499H}). 

\section{Conclusion}
\label{sec:conclusion}
We report flare ribbon fine structures observed during a GOES C2-class flare using high-resolution, multi-line spectropolarimetric observations from DKIST. The high spatial and spectral resolution of the ViSP data enabled us to resolve subarcsecond structures within the ribbons. In the red wing of the \ion{Ca}{2} line, we identified compact, bright, and spatially quasi-periodic ribbon blobs exhibiting asymmetric, double-peaked profiles and enhanced intensity. Multi-line non-LTE inversions of the ViSP spectra revealed that these blobs are associated with localized heating (up to $\sim$8~kK). Moreover, the LOS velocity in the ribbons showed the presence of both upflows and downflows at two heights in the chromosphere. Our findings support the interpretation that ribbon fine structures may result from localized energy release associated with magnetic reconnection, potentially linked to tearing-mode instabilities in the flare current sheet. 
Our work underscores the importance of high-temporal and high-spatial resolution observations for probing the dynamic nature of ribbon fine structures in flares.

\begin{acknowledgements}
We thank the anonymous referee for their constructive comments and suggestions, which have significantly improved this paper. 
We thank the US government for providing the funding that made this research possible. We acknowledge support from NASA ECIP NNH18ZDA001N and NSF CAREER SPVKK1RC2MZ3.
The research reported herein is based in part on data collected with the Daniel K. Inouye Solar Telescope (DKIST) a facility of the U.S. National Science Foundation.  DKIST is operated by the National Solar Observatory under a cooperative agreement with the Association of Universities for Research in Astronomy, Inc. DKIST is located on land of spiritual and cultural significance to Native Hawaiian people. The use of this important site to further scientific knowledge is done with appreciation and respect. This research has made use of NASA’s Astrophysics Data System. We acknowledge the community effort devoted to the development of the following open-source packages that were used in this work: NumPy (\url{numpy.org}), matplotlib (\url{matplotlib.org}) and SunPy (\url{sunpy.org}).

\end{acknowledgements}

\bibliographystyle{aa}
\bibliography{new-ref}

\end{document}